\documentstyle[12pt,aasms4]{article}

\lefthead{Allende Prieto, Garcia Lopez and Trujillo Bueno}
\righthead{stellar line asymmetries}
\begin{document}

\title{The Limited Influence of Pressure Gradients on Late-type Stellar Line Asymmetries.}

\author{Carlos Allende Prieto, Ram\'on J. Garc\'\i a L\'opez, and Javier Trujillo Bueno}

\affil{Instituto de Astrof\'\i sica de Canarias \\ E-38200, La Laguna, Tenerife,  \\  SPAIN}

\authoraddr{IAC, E-38200, La Laguna, Tenerife,  SPAIN}

\authoremail{callende@iac.es, rgl@iac.es, and jtb@iac.es}

\begin{abstract}

Line asymmetries and shifts are a powerful tool for studying velocity
fields in the stellar photospheres. Other effects, however, could also
generate asymmetries blurring the information of the velocity
patterns.  We have studied the shifts and asymmetries induced in the
profiles of spectral lines  by pressure effects. The best theoretical
and experimental data on line broadening and shifts caused by
collisions with atomic hydrogen were used to analyze the Na\,{\sc i} D
and three Ca\,{\sc i} lines. Line bisectors of synthetic spectra
computed with accurate data for the Na\,{\sc i} and Ca\,{\sc i} lines
are compared with very high resolution high signal-to-noise ratio solar
spectra and indicate that pressure broadening reproduces the wings of
the observed lines, but pressure shifts  introduce neither asymmetries
nor shifts comparable to the observed ones.

\end{abstract}

\section{Introduction} 
\label{sec1}

In the study of stellar atmospheres, new insight is coming from the very high resolution spectroscopic techniques. Important physical information is contained in the spectral line asymmetries. They are supposed to be due to velocity fields in the line formation region and, in particular for late-type stars, they are mainly associated with solar-like granular motions (\markcite{gray80}Gray 1980; \markcite{dravins81}Dravins, Lindegren, \& Norlund 1981). 

In metal-poor stars, with atmospheres in which the lower opacity could be responsible for an enhanced convective influence on the line formation, stronger deviations from symmetry are expected. \markcite{viena}Allende Prieto et al. (1995) observed with very high spectral resolution ($\lambda /\Delta\lambda\sim 170000$) a set of isolated lines in the star HD140283, with a metallicity $\sim 500$ times lower than that of the Sun. They found that the bisectors of the lines show the same kind of ``C'' shape observed in late-type metal-rich stars but with a more pronounced shift to the red in the wings. Those observations are part of a wider project aimed at better understanding the atmospheres of metal-poor stars.

Velocity gradients and granulation patterns are, however, not the only potential factors that can produce  spectral line asymmetries. Blends with other lines may occur, although they will be less important for metal-poor stars than for stars with higher metallicity. Isotopic shifts and hyperfine structure will split  the line profile in different components, and introduce asymmetry (\markcite{kuruczisot93}Kurucz 1993). 

On the other hand, it is well-known that collisions between an emitting/absorbing atom in a plasma and the surrounding particles induce not only a symmetric broadening (\markcite{victor32}Weisskopf 1932), but also a shift in the line  emission/absorption coefficient (\markcite{lindholm42}Lindholm 1942,\markcite{lindholm45}1945; \markcite{foley46}Foley 1946). That displacement is proportional to the density of perturbers. The quantum-mechanical reformulations (\markcite{anderson52}Anderson 1952, \markcite{griem62}Griem et al. 1962) do not qualitatively change the classical result. When applied to stellar atmospheres, since the line formation takes place along a density stratified region, asymmetries and a net shift in the central wavelength of the emergent profile will appear. These line shifts were suggested to be the origin for the solar limb effect (\markcite{spitzer50}Spitzer 1950; \markcite{hart74}Hart 1974), although later the convective blue-shift, resulting from the correlation between intensity and velocity fields in the solar photosphere, became a better candidate (\markcite{BeckVeg79}Beckers \& de Vegvar 1978; \markcite{BeckNelson78}Beckers \& Nelson 1978). In these previous studies, making use of the Lennard-Jones potential (\markcite{hindmarsh67}Hindmarsh, Petford, \& Smith 1967) and the classical impact theory, pressure shifts in the line absorption profile were estimated at a representative atmospheric height only. This provides an indication of the mean displacement of the line but does not give information about possible asymmetries in its shape. The asymmetries will depend on the variation of the pressure shifts from the top to the bottom of the line formation region. \markcite{vince87}Vince \& Dimitrijevi\'{c} (1987) analyzed the influence of pressure shifts in the spectral line formation making use of the Roueff theory (\markcite{roueff70}Roueff 1970), and found that the effect might be responsible for up to $20-30$\% of the asymmetry observed in the Na\,{\sc i} $\lambda$ 6160.8 \AA\ solar line.  

For a better understanding of the behaviour of line asymmetries in the atmospheres of late-type stars, we analyze in this paper the potential asymmetries present in the line profile due to collisional shift gradients. The study is carried out for a few atomic lines which have been the object of the best theoretical and experimental studies on line broadening and shifts caused by collisions with atomic hydrogen.

We have concentrated mainly on the Sun, for which observed spectra with the highest resolution and signal-to-noise ratio are available. In section 2 we estimate the asymmetries in the spectral profiles of Na\,{\sc i} and Ca\,{\sc i} lines emerging from the solar atmosphere. We also estimate the shifts for a given Ca\,{\sc i} line in a metal-poor star and compare them with our observations. The main conclusions are summarized in section 3.

\section{Pressure-induced line asymmetries}
\label{sec2}

To study the importance of pressure effects as asymmetry generators in solar spectral lines, we integrated the transfer equation assuming the solar photosphere to be plane-parallel and horizontally homogeneous, in hydrostatic equilibrium, and  in local thermodynamic equilibrium (LTE). Radiative damping was neglected. The collisional damping and the velocity broadening are assumed to be statistically independent, so the line absorption profile takes the familiar Voigt shape. We introduced a shift of the line absorption profile, proportional to the hydrogen density,  at every depth. A shift of the Lorentzian collisional profile is equivalent to a shift of the Voigt profile. The dependence of width and shift  on temperature is assumed to be a power law. We fitted by least squares the corresponding theoretical data for the line, independently for the line width and shift, and the resultant slopes are shown in Table 1. For this test, we used the Na\,{\sc i} and Ca\,{\sc i} lines listed in Table 2, for which accurate theoretical data are available. The \markcite{holmu74}Holweger-M\"uller (1974; HM) solar model atmosphere was employed. 

The line asymmetry was quantified by the line bisector  (\markcite{kulander66}Kulander \& Jefferies 1966; \markcite{dravins81}Dravins et al. 1981). For comparison purposes, the asymmetry due to the wavelength variation of the source function and the continuum absorption coefficient across the line were taken into account. To compare with observed spectra and to estimate the real significance of pressure-induced asymmetries and shifts, line bisectors were measured, when possible, for the considered lines in the \markcite{kuruczatlas84}Kurucz et al. (1984) solar flux atlas. 

Several factors were taken into account in choosing the lines to study. Laboratory measurements or accurate theoretical results on collisional shifts and widths by atomic hydrogen are available only for the following cases: the Na D lines, the Ca\,{\sc i} $4s4p-4s5s$ triplet, the Ca\,{\sc ii} H and K lines and the Mg\,{\sc ii} h and k lines. Furthermore, it has been recognized that non-LTE effects are negligible in the wings of the Na\,{\sc i} and Ca\,{\sc i} lines (\markcite{gehren75}Gehren 1975, \markcite{drake91}Drake \& Smith 1991), but the Ca\,{\sc ii} H and K lines suffer strong deviations from LTE even in the far wings (\markcite{castelli88}Castelli et al. 1988).  The magnesium doublet lies in the ultraviolet, outside the boundaries of the \markcite{kuruczatlas84}Kurucz et al. (1984) solar atlas. For these reasons we concentrated our study on the sodium D lines and on the 4s4p-4s5s multiplet of neutral calcium. For sodium and calcium, isotopic and hyperfine splitting are not significant, eliminating some additional uncertainties.  The basic data for the Na D and Ca\,{\sc i} lines are listed in Table 2. 

\subsection{Na D lines}

Na D lines have been the subject of the only laboratory experiments using atomic hydrogen. \markcite{omara86}O'Mara (1986) discusses the two existing results, giving preference to the measurements by \markcite{lemaire85}Lemaire, Chotin, \& Rostas (1985). The  theoretical calculations of  \markcite{monteiro85}Monteiro, Dickinson, \& Lewis (1985), which do not distinguish between the two lines, are in good agreement with the laboratory measurements for broadening, while there is a factor two of difference in the shifts. Experimental results of \markcite{lemaire85}Lemaire et al. (1985) and several theoretical values are compiled in Table 3. 

\markcite{smith85}Smith et al. (1985) found a fit to the wings of the solar features with the collisional damping constant determined by \markcite{monteiro85}Monteiro et al. (1985). Figure 1 shows the comparison of two synthetic spectra, computed with solar sodium abundance for the experimental and theoretical line broadening values of \markcite{lemaire85}Lemaire et al. (1985) and \markcite{monteiro85}Monteiro et al. (1985), respectively, and the solar atlas of \markcite{kuruczatlas84}Kurucz et al. (1984). Also plotted, as comparison, are synthetic spectra computed for the Na D lines assuming the Lennard-Jones interaction potential. Although no special efforts were devoted to find parameters which best fit the solar lines, and there is an uncertainty with respect to location of the continuum level, it can be seen in Fig. 1 that the experimental value of \markcite{lemaire85} Lemaire et al. (1985) is close to the observed spectrum. The cores of these lines are affected by deviations from  LTE (\markcite{covino93}Covino et al. 1993) and are not reproduced by the synthetic spectra.

Line bisectors were determined for the synthetic profiles and are shown in Figure 2. Dashed lines correspond to the experimental (bluer bisector) and theoretical (redder bisector) values for the pressure shift. The wavelength dependences of the source function (dotted-dashed line) and that of the continuum absorption coefficient (dotted line) translate into a bisector bent to the blue, while the pressure  induces a redwards asymmetry. The three effects are of comparable size for these lines, and an order of magnitude smaller than the asymmetries observed in the solar lines. Taking into account the three effects, the resulting bisector has a reversed ``C'' shape (solid line) including the experimental (left-hand curve) or theoretical (right-hand curve) values for the pressure shifts.

Unfortunately, line bisectors cannot be measured reliably in the solar spectrum for these lines; other solar and telluric lines are blended with the wings of the Na D lines. Nonetheless, the predicted asymmetries are likely much smaller than those of the observed lines.  

The synthetic spectra show a line shift of less than 1 m/s, as anticipated from the small ratio of the  pressure shift ($d$) to width ($w$) ratio which, following \markcite{lemaire85}Lemaire et al. (1985), is as small as $d/w = -0.037$, being $w$ the half width at half maximum. \markcite{pierce73}Pierce \& Breckinridge (1973) list an accurate (error of the mean: 31 m/s) measurement of the solar wavelength of the Na D$_{1}$ line, which compared with its laboratory wavelength and corrected for the gravitational redshift, corresponds to a 220 m/s blueshift. This shift must be of convective origin.

\subsection{Ca\,{\sc i} $4s4p-4s5s$ multiplet}

No experimental measurements of line shift and broadening by collisions with hydrogen have been carried out for this triplet. However, detailed calculations have been performed by \markcite{spielfiedel91}Spielfiedel et al. (1991), and the predicted broadening for the $\lambda$ 6162 \AA\ nicely agrees with that empirically measured in the Sun by \markcite{smith86}Smith et al. (1986). Theoretical and empirical values for this line are listed in Table 4.

We fitted the wings of the solar Ca\,{\sc i} lines with the solar calcium abundance given by \markcite{smith81}Smith (1981), $\log {\rm N(Ca)}=6.36$ on the scale in which the hydrogen abundance is 12. Theoretical  and empirical values for the damping constant provide synthetic wing profiles close to those observed in the solar atlas  for the Ca\,{\sc i} $\lambda$ 6162 \AA\ line (see Figure 3). The differences between the two synthetic profiles do not modify the conclusions regarding the asymmetries. The dominant effect is the pressure gradient, while the spectral dependence of the source function and the continuum absorption coefficient make a negligible contribution. 

\markcite{lambert93}Lambert (1993) addresses the question of whether the differences in the collisional widths calculated by \markcite{spielfiedel91}Spielfiedel et al. (1991) among the three lines of the triplet would be reflected in the stellar spectra.  Figure 4 shows how there are real differences in the solar line widths corresponding to the predictions of the theoretical calculations.  It has been assumed that the continuum normalization of the \markcite{kuruczatlas84}Kurucz et al. (1984) solar atlas is good enough for this comparison, and plotting a longer spectral range than that shown in the figure that hypothesis seems to be correct. The synthetic line profiles  were vertically shifted to fit the corresponding solar atlas lines, and it can be appreciated that their continua are not coincident.

In accordance with the pressure shift to width ratios  $(d/w)$ in the line absorption profile in the range 0.07--0.1, the shifts measured in the emerging line profile were (to the blue) in the range 0.5--2.5 m/s for the three lines. Comparing the solar wavelengths published by \markcite{pierce73}Pierce \& Breckinridge (1973) with those ones measured in the laboratory, and taking into account the redshift due to the difference in solar and terrestrial masses, we found that the solar calcium lines are shifted from $-130 $ to $-350$ m/s, with an estimated error around 100 m/s. These observed shifts, arising presumably from convective motions, are two orders of magnitude greater that those predicted here.

The triplet profiles are almost immaculate in the solar spectrum, allowing the measurement of the bisectors  to within $20$\% of the continuum, or even higher. Computed and observed bisectors for the three lines in the Sun are plotted in Figure 5. Their comparison clearly reveals that the pressure shifts are unimportant contributors to the observed asymmetries. Therefore, at least with respect to the three considered mechanisms, the study of velocity fields in the solar photosphere would be direct from the asymmetries observed in these Ca\,{\sc i} lines. Hotter mean-sequence stars have lower gas pressure and hydrogen density in their photospheres, and even smaller pressure effects. Cooler dwarfs have higher density and higher surface gravity, increasing the magnitude of the effects. We have compared predicted pressure-induced asymmetries for the $\lambda$ 6162 \AA\ line in the Sun and in a K5 V type star (computed using a \markcite{kurucz92}Kurucz 1992 model atmosphere). Although the asymmetry is much more important in the later-type star, it is well below the strength of the solar observed asymmetry for that line, and observations point to stronger velocity fields in those stars than in the Sun (\markcite{gray82}Gray 1982; \markcite{dravins87}Dravins 1987).

To test the use of line asymmetries for studying velocity fields in metal-poor stars, we have extended the analysis of the Ca\,{\sc i} $\lambda$ 6162 \AA\ line to the very metal-poor star HD140283 ([Fe/H]$=-2.7$; where [Fe/H]$=\log ({\rm Fe/H})_* -\log ({\rm Fe/H})_\odot$). This line is included within the spectral range covered by the very high-resolution observations of the star performed at McDonald Observatory (Texas) using the 2d-coud\'e spectrograph (\markcite{tull95}Tull et al. 1995) at the 2.7m telescope. The metal deficient atmosphere, resulting from the lack of donors compared with the sun, has a lower electron pressure. Also lower temperature, opacity, and gas pressure. This  gives place to a very significant reduction of the role of the pressure-induced asymmetries. The observed bisector for this line behaves like the neutral iron lines as shown by \markcite{viena}Allende Prieto et al. (1995), with a redward asymmetry up to 400 m/s. The pressure-induced  bluewards asymmetry is smaller than 1 m/s. Enhanced convection due to the lower opacity is likely the main responsible for the line asymmetries in metal-poor stars.

\section{Conclusions}
\label{sec4}

We have analyzed carefully the asymmetries associated with pressure effects in a limited sample of spectral lines. We used the most accurate experimental and theoretical data available for Na D and three Ca\,{\sc i} lines. All of the computed synthetic spectra were compared with their corresponding counterparts in the solar spectrum. The predictions for the Ca\,{\sc i} $\lambda$ 6162 \AA\ line were also compared with a high resolution spectrum of a very metal-poor star.
 
The conclusions of this work can be summarized as follows:

\begin{itemize}

\item Pressure shifts are shown not to be an important contributor to the asymmetry observed in several solar spectral lines corresponding to neutral elements, but the conclusion cannot be generally extended.

\item The differences in the collisional widths theoretically predicted by Spielfiedel et al. (1991) for the three lines of the Ca\,{\sc i} $4s4p-4s5s$ multiplet are reflected in the solar spectrum.

\item The effect of the collisions are negligible when analyzing line asymmetries in very metal-poor stars.

\end{itemize}

The comparison made by \markcite{anstee95}Anstee \& O'Mara (1995) between their results on line broadening and the solar spectrum give confidence to their analysis. The extension of those calculations to shift cross-sections would be very important. In particular, the application of the detailed molecular-like potential methods to many lines of astrophysical interest. This would allow to analyze the expected effects in different lines, and would help to prevent errors in the studies of velocity fields using line asymmetries. This should be carried out by promoting collaborations between theoretical and experimental physicists and astrophysicists.

\acknowledgements

It is a pleasure to acknowledge the following people for their help with the analyses described in this work: R. I. Kostik, D. L. Lambert, T. Meylan, F. de Pablos, J. S\'anchez Almeida, N. G. Shchukina, and G. Smith. This research has made use of the Vienna Atomic Line Data-Base (VALD), and was partially supported by the Spanish DGICYT under projects PB92-0434-c02-01 and PB91-0530.

\clearpage


\begin{table*}
 \begin{center}
   TABLE 1\linebreak\bigskip
   {\sc Exponents for the width and shift dependences on temperature}\bigskip

  \begin{tabular}{cccc}
   \tableline
   \tableline
   & & & \\
   Spectral line & Width & Shift & Reference\\ 
   & & & \\
   \tableline
   & & & \\
   Na D                       &   0.41 &  0.22   & Monteiro et al. (1985)\\
   Ca\,{\sc i} $\lambda$ 6103 &   0.39 &  0.80   & Spielfiedel et al.  (1991) \\ 
   Ca\,{\sc i} $\lambda$ 6122 &   0.40 &  1.16   &  ~~~" ~~~ " \\ 
   Ca\,{\sc i} $\lambda$ 6162 &   0.40 &  1.20   &  ~~~" ~~~ " \\
   \tableline 
\end{tabular}
\end{center}
\tablenum{1}
\end{table*}

\clearpage


\begin{table*}
 \begin{center}
   TABLE 2\linebreak\bigskip
   {\sc Basic data for the lines used in the computations}\bigskip

  \begin{tabular}{cccl}
   \tableline
   \tableline
   & & &  \\
   Element  &  Wavelength & Ex. Potential & ~~$\log gf$\tablenotemark{a} \\
            &  (\AA)      & (eV)          &           \\
   & & &  \\
   \tableline
   & & &  \\
Na\,{\sc i} &  5889.95    & 0.000   & $+$0.117 \\
Na\,{\sc i} &  5895.92    & 0.000  & $-$0.184 \\
Ca\,{\sc i} &  6102.72    & 1.879  & $-$0.793 \\ 
Ca\,{\sc i} &  6122.22    & 1.886  & $-$0.316 \\  
Ca\,{\sc i} &  6162.17    & 1.890  & $-$0.097 \\ 
\tableline
\end{tabular}
\end{center}
\tablenum{2}

\tablenotetext{a}{Oscillator strenghts ($\log gf$) come from  Wiese et al. (1969) and Smith \& O'Neill (1975) for the sodium and calcium lines, respectively.}

\end{table*}
\clearpage

\begin{table*}
 \begin{center}
   TABLE 3\linebreak\bigskip
   {\sc Widths and shifts for the {\rm Na}\,{i} D lines}\bigskip

  \begin{tabular}{cccc}
   \tableline
   \tableline
   & & & \\
 Reference & Analysis &  $w/N_H$                               & $d/N_H$ \\ 
           &          &\multicolumn{2}{c}{($\times 10^{-9}$ rad cm$^3$ s$^{-1}$)} \\
   & & & \\
\tableline
   & & & \\
   Lemaire et al. (1985)  &  Experimental & $13.9\pm 1.9$ & $-0.52\pm 0.16$\\
   Lewis et al.   (1971)  &  Theoretical  &   8.3         & $-0.59$  \\
   Monteiro et al. (1985) &  Theoretical  &  10.6         & $-1.1$  \\
   Anstee \& O'Mara (1995)&  Theoretical  &  11.67        & \dots\nonumber\\
\tableline
\end{tabular}
\end{center}
\tablenum{3}
\tablecomments{$w/N_H$ and $d/N_H$ represent the half width at half maximum and the shift (both in angular frecuency units) over the hydrogen number density, respectively.}
\end{table*}

\clearpage


\begin{table*}
 \begin{center}
   TABLE 4\linebreak\bigskip
   {\sc Widths and shifts for the {\rm Ca}\,{i} $\lambda$ 6162 \AA\ line}\bigskip

  \begin{tabular}{cccc}
   \tableline
   \tableline
   & & & \\
 Reference & Analysis &  $w/N_H$                               & $d/N_H$ \\ 
           &          &\multicolumn{2}{c}{($\times 10^{-9}$ rad cm$^3$ s$^{-1}$)}\\
   & & & \\
\tableline
   & & & \\
   O'Neill \& Smith (1980)   & Solar Empirical & $24\pm 6$ &  \dots\nonumber \\
   Smith et al. (1986)       & Solar Empirical & $30\pm 2$ &  \dots\nonumber \\
   Spielfiedel et al. (1991) & Theoretical     &  25.8     &   2.23 \\ 
   Anstee \& O'Mara (1995)   & Theoretical     &  24.9     &  \dots\nonumber \\
\tableline 
\end{tabular}
\end{center}
\end{table*}

\clearpage

\figcaption{Sodium D$_{1}$ (a) and D$_{2}$ (b) line profiles. Different theoretical and experimental values for the collisional line width are compared with the solar atlas of Kurucz et al. (1984).}

\figcaption{Intrinsic (dotted-dashed line: source function; dotted line: continuum absorption coefficient) and pressure-induced (dashed curves) line asymmetries of the sodium D lines in the solar atmosphere. Theoretical (redder dashed  line) and laboratory measurements (bluer dashed line) of the pressure width and shifts are compared.  The solid line shows the total effect.}

\figcaption{Synthetic profiles of the Ca\,{\sc i} $\lambda$ 6162 \AA\ line, calculated using theoretical and empirical determinations of the  collisional widths,  are compared with the solar spectrum (see text for details).}

\figcaption{Upper Figure: differences between the profiles of the three Ca\,{\sc i} $4s4p-4s5s$  multiplet solar lines. Lower Figure: profiles synthesized using the theoretical values of Spielfiedel et al. (1991) for the collisional line widths. Continuum levels have been individually shifted to match the solar spectrum.}

\figcaption{Comparison between pressure-induced bisectors computed using Spielfiedel et al. (1991) data (dotted line) and those measured in the solar lines (solid line) for the Ca\,{\sc i}   $4s4p-4s5s$ multiplet.}  

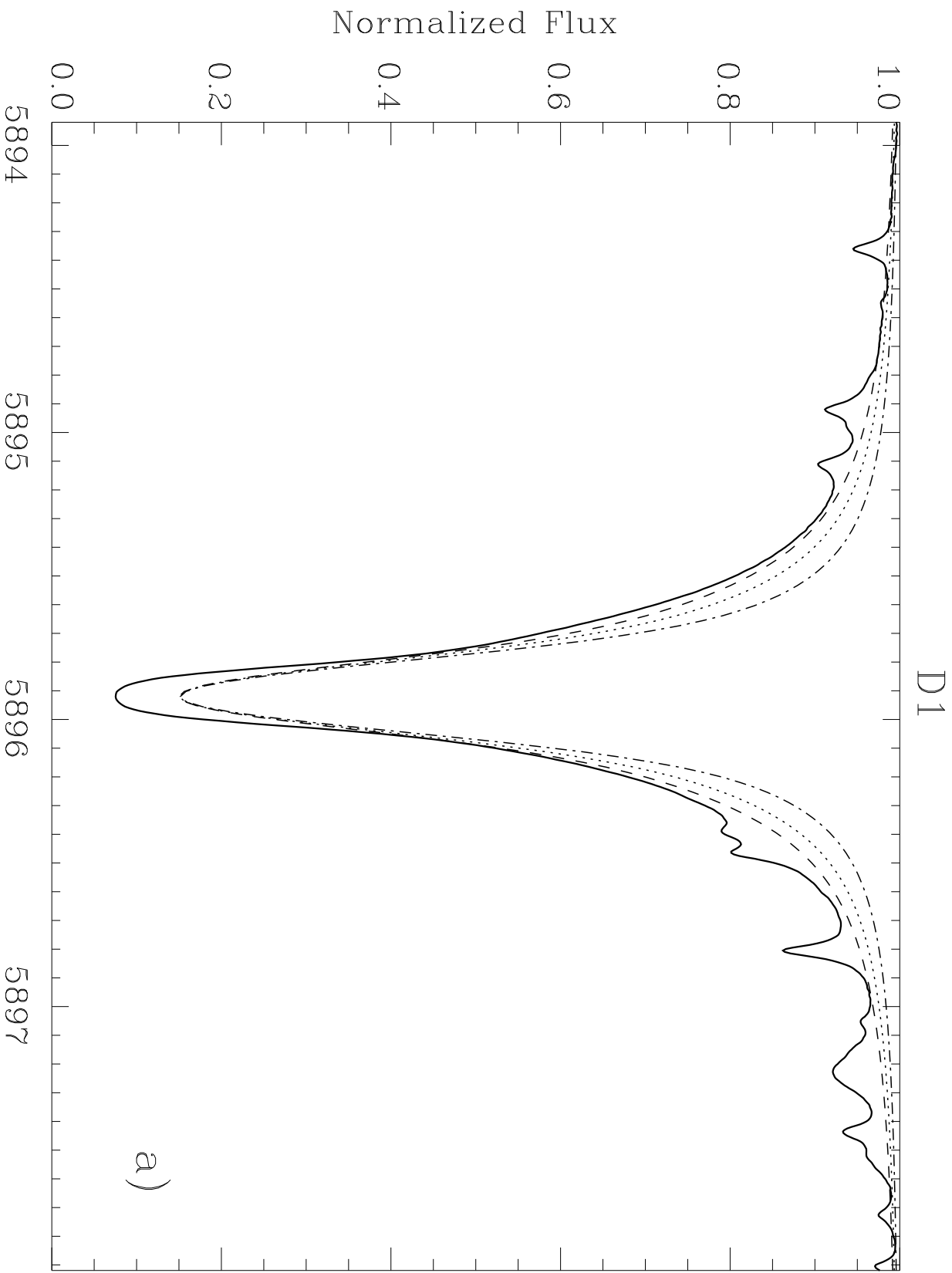
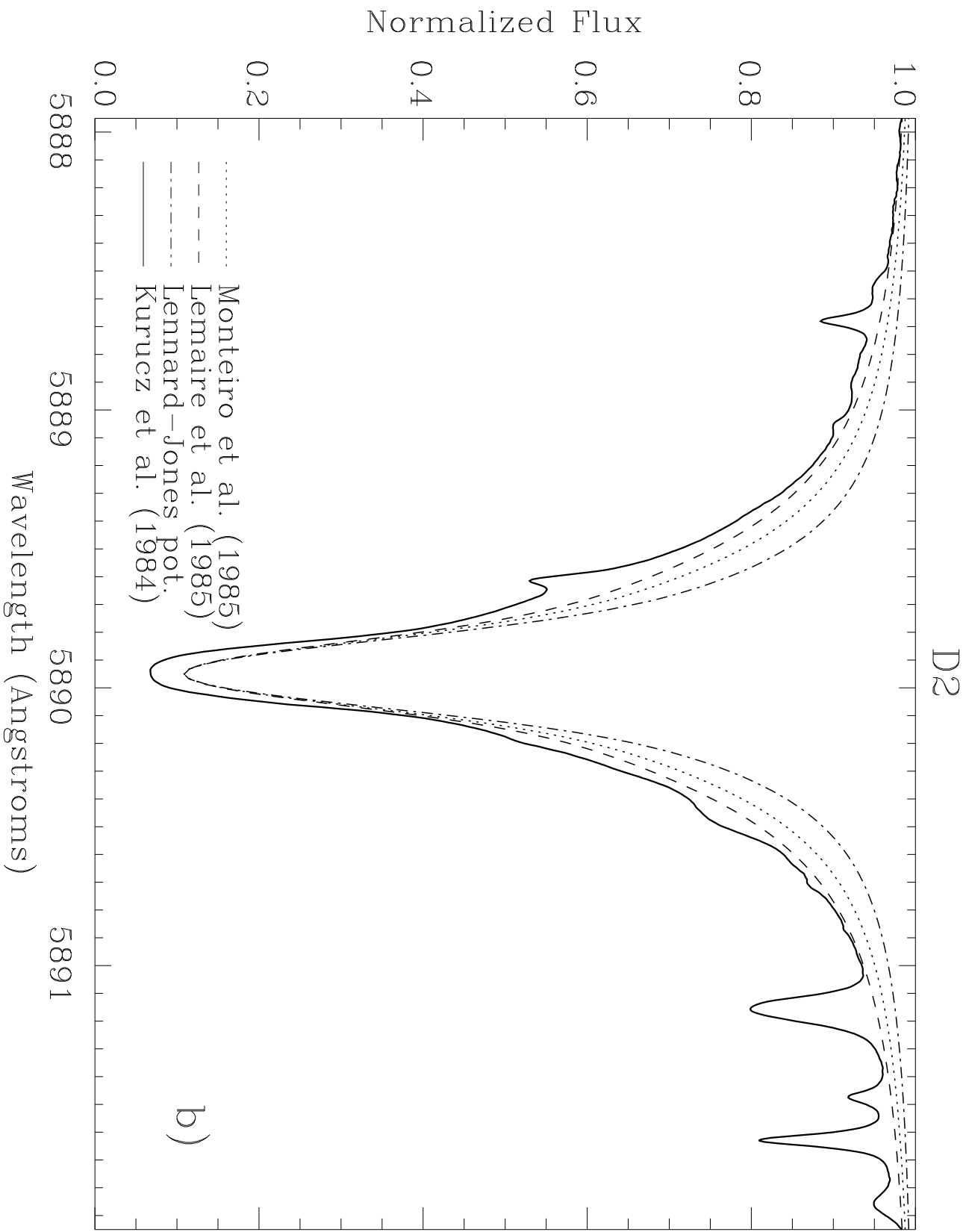
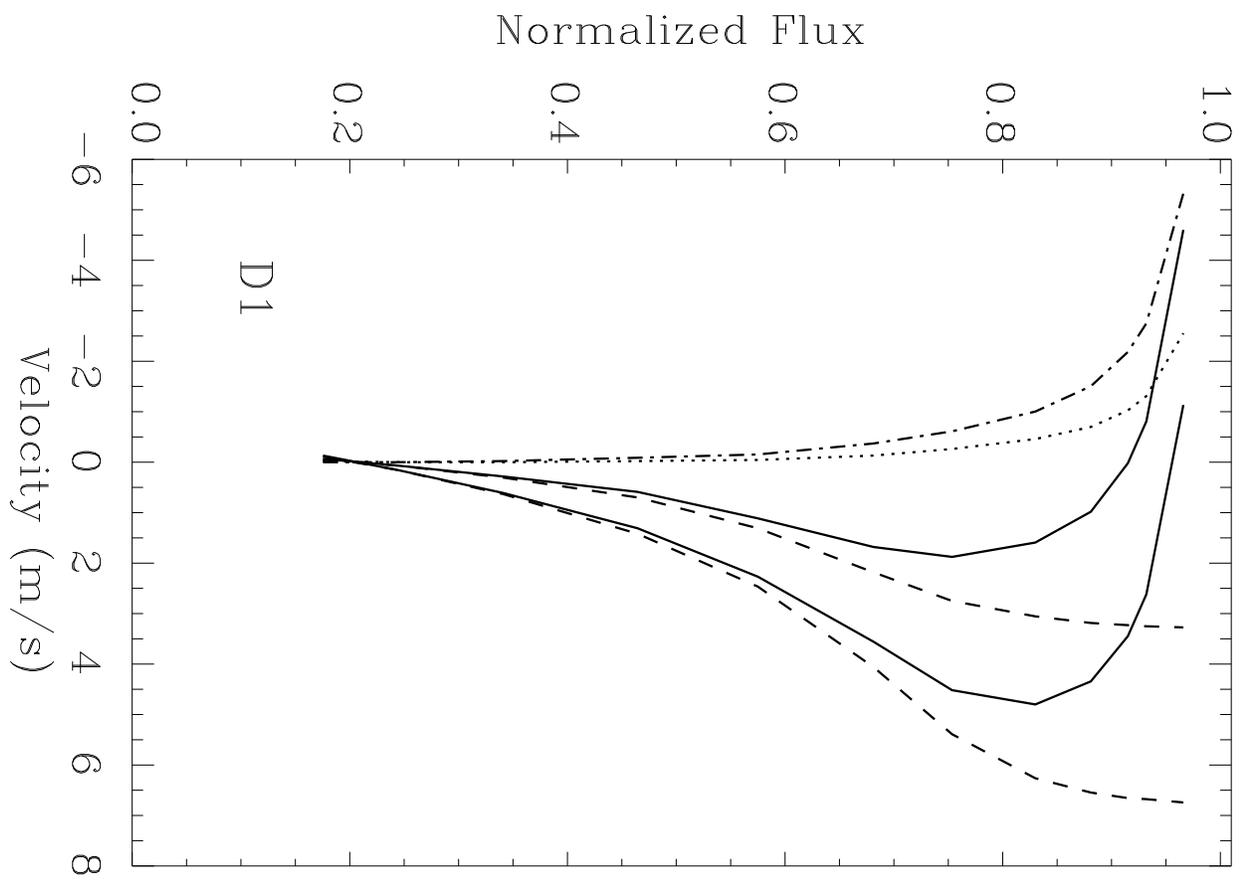
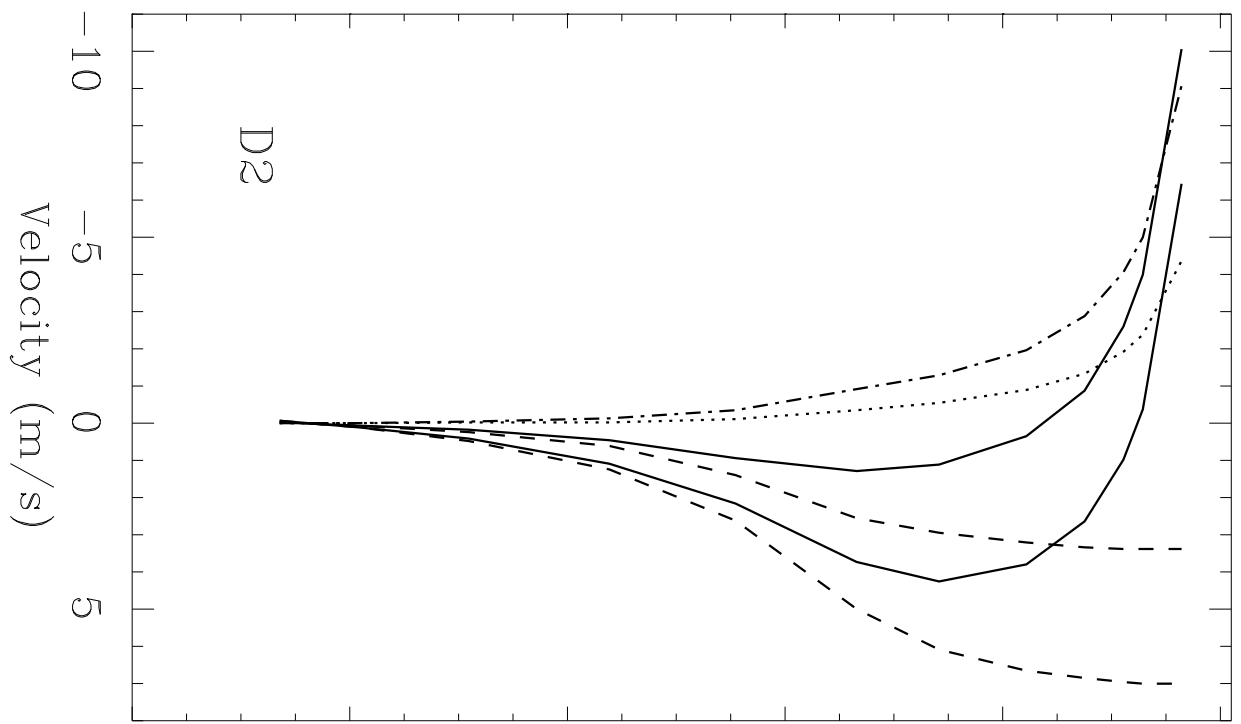
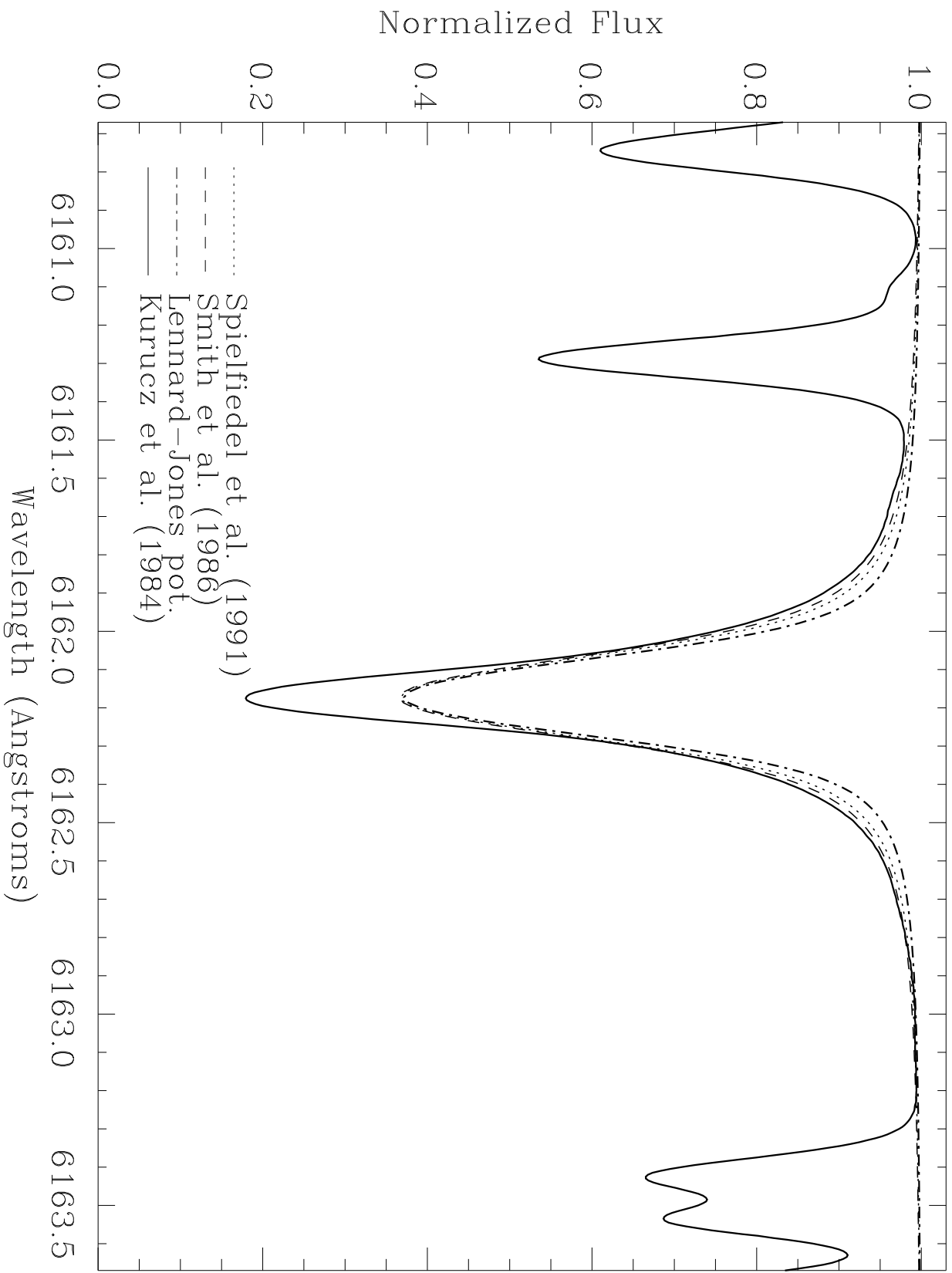
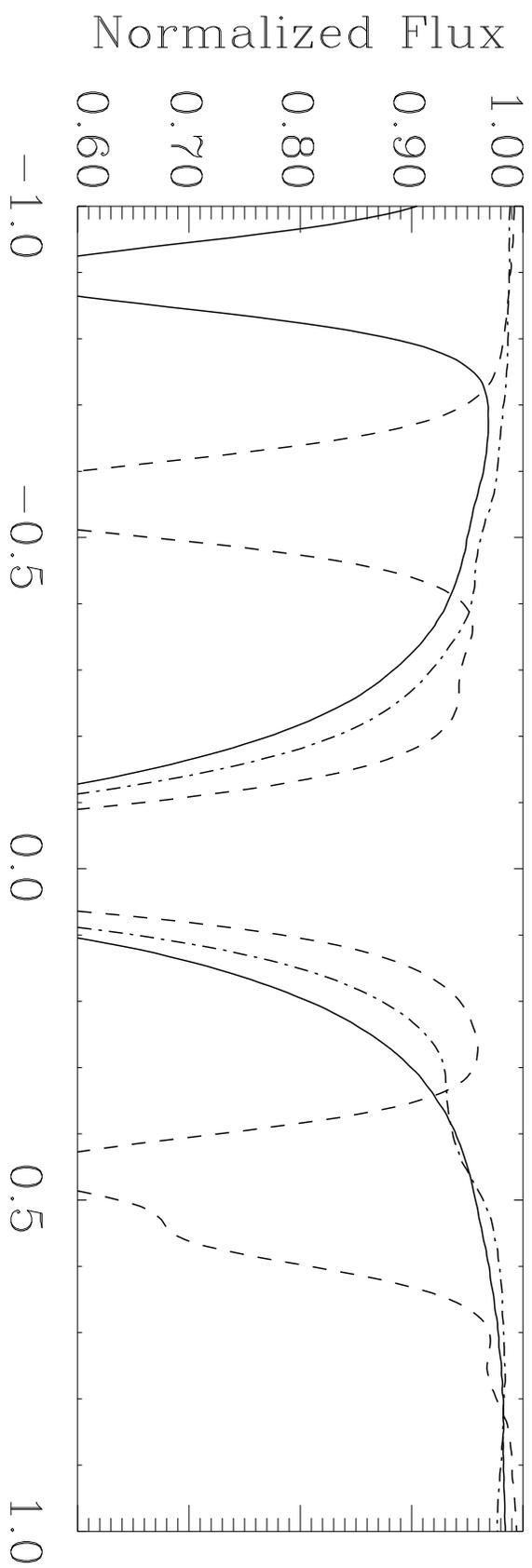
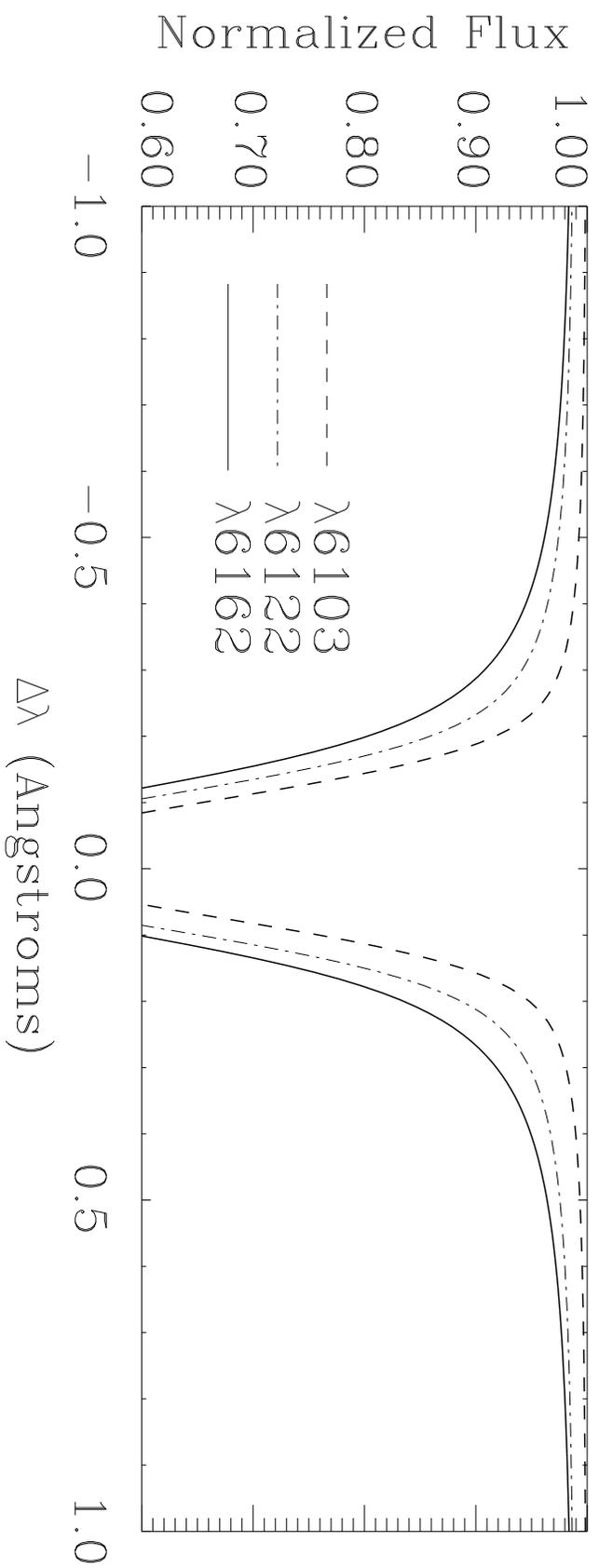
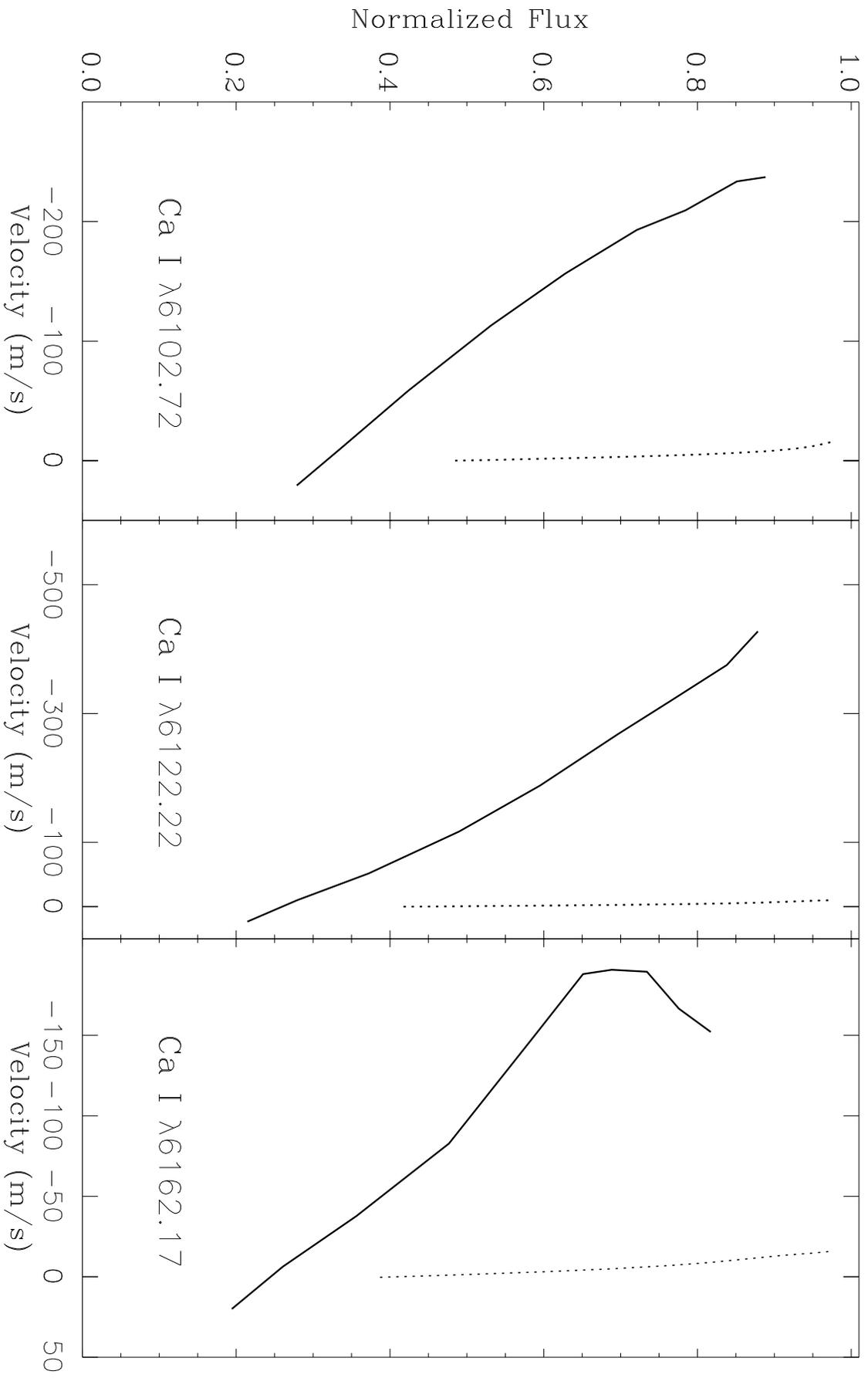
\end{document}